\documentclass[10pt,a4paper]{article}

\usepackage[latin1]{inputenc}
\usepackage{times}
\usepackage{graphicx}
\usepackage{amsmath}
\usepackage{amsfonts}
\usepackage[latin1]{inputenc}
\usepackage[T1]{fontenc}
\usepackage{verbatim}
\usepackage{latexsym}
\usepackage{natbib}

\usepackage{hyperref}
\usepackage{url}

\hypersetup{backref,colorlinks=true,citecolor=blue}

\title{Determining the wavelength of Langmuir wave packets at the Earth's bow shock}

\author{V.~V. Krasnoselskikh$^1$, T. Dudok de Wit$^1$, S. D. Bale$^2$ \\
\small $^1 $ Laboratoire de Physique et Chimie de l'Environnement et de l'Espace,\\
\small UMR 6115 CNRS and University of Orl\'eans, 3A avenue de la Recherche Scientifique, 45071 Orl\'eans, France\\
\small $^2 $ Physics Department and Space Sciences Laboratory, University of California, Berkeley, CA 94720-7450, USA}

\date{\normalsize \textit{This article has been published in Annales Geophysicae 29 (2011) 613-617. DOI:10.5194/angeo-29-613-2011}}

\begin{document}
\maketitle

\sloppy

\begin{abstract}
The propagation of Langmuir waves in plasmas is known to be sensitive to
density fluctuations. Such fluctuations may lead to the coexistence of wave
pairs that have almost opposite wave-numbers in the vicinity of their
reflection points. Using high frequency electric field measurements from the
WIND satellite, we determine for the first time the wavelength of intense
Langmuir wave packets that are generated upstream of the Earth's electron
foreshock by energetic electron beams. Surprisingly, the wavelength is found
to be 2 to 3 times larger than the value expected from standard theory. These
values are consistent with the presence of strong inhomogeneities in the
solar wind plasma rather than with the effect of weak beam instabilities.
\end{abstract}

\section{Introduction}
\label{sec_intro}

The theory of beam-plasma interactions in homogeneous plasmas is one of the
pillars of plasma physics, and yet it fails to properly describe some
important physical phenomena observed, such as the generation of type III
solar radio bursts or waves registered in the electron foreshock region,
upstream of the Earth's bow shock \citep{ginzburg70, robinson97}. The major
reason for this discrepancy between observations and the theory of
quasilinear, weak or strong turbulence is supposed to be related to the
existence of strong density fluctuations that are not associated with the
beams. When these fluctuations are large enough, Langmuir waves can be
trapped or reflected and in both cases the wave-vector $\vec{k}_{1}$ of the
reflected wave becomes approximately equal and opposite to that of the
incident wave $\vec{k}_{0}\simeq -\vec{k}_{1}$. Such strong density
fluctuations are not resolved by existing particle detectors. However, they
provide a useful means for inferring the wavelength, which is difficult to
measure directly in space plasmas.

\begin{figure}[t]
\vspace*{2mm}
\center
\includegraphics[width=9cm]{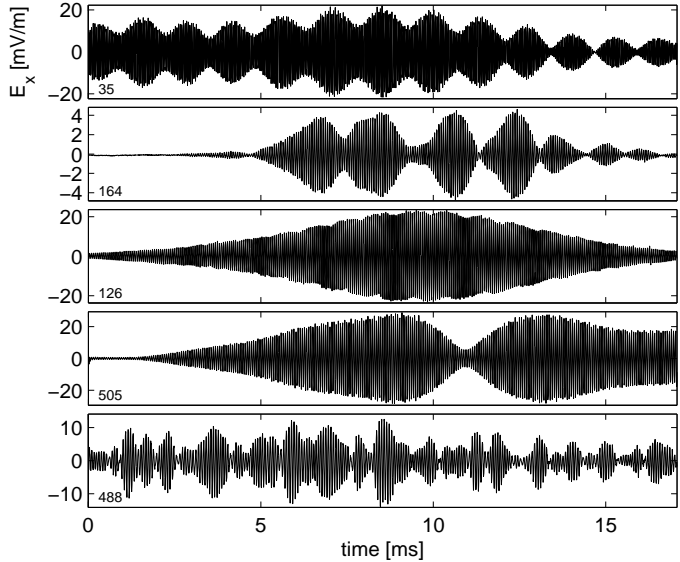}
\caption{Five typical coherent Langmuir wave packets observed by WIND.
Only the $E_{\rm x}$ electric field component is shown. The bottom plot shows a
marginal case in which the wave-packet is barely coherent.}
\label{fig_example}
\vspace*{-2mm}
\end{figure}

The WAVES instrument \citep{bougeret95} onboard the WIND satellite routinely
measures Langmuir waves at high time resolution (up to 120\,000 samples per
second) in the solar wind. When two large amplitude and counter-propagating
waves coexist, a monochromatic wave packet with a slowly varying envelope is
produced. Some typical examples are shown in Fig.~\ref{fig_example}. Here,
for the first time, we show how the wavelength of these waves can be directly
estimated from a polarization analysis of two components of the electric
field. Interestingly, the measured value disagrees with estimates derived
from standard beam-plasma theory in homogeneous plasmas but is consistent
with estimates derived for strongly inhomogeneous plasmas.

\section{The effect of density fluctuations}
\label{sec_effect}

Density fluctuations in the solar wind plasma have been studied for many
years \citep{neugebauer75, celnikier87} and exhibit a power law over many
decades. The smallest scales are difficult to measure by lack of time
resolution and counting statistics of standard particle instruments.
Recently, \citet{ergun08} observed the interaction of density fluctuations
with Langmuir waves, and described them as eigenmodes trapped inside density
depletions. In such strongly inhomogeneous plasmas, wave propagation is
strongly affected by density fluctuations. As a consequence, the dynamics of
the instability departs quite substantially from that observed in homogeneous
plasmas \citep{nishikawa76, muschietti85, krasno07}. Such density
inhomogeneities can cause high frequency electrostatic waves to be converted
into electromagnetic waves \citep{bale98,kellogg99}. They also affect the
statistics of wave amplitudes \citep{cairns00,krasno07} and the presence of
sufficiently large inhomogeneities eventually leads to wave reflection
\citep{ginzburg70, budden88}.

%f1

Here, we focus on the dynamics of Langmuir waves that are generated near the
electron foreshock edge region by electron beams with a typical energy of
1--3\,keV \citep{filbert79}. These waves have a narrowband spectrum and their
characteristic frequency is close to the electron plasma frequency. The
presence of relatively large but slowly varying (with respect to the
wavelength) density fluctuations, with
\begin{equation}
\delta n/n_{0}\gg \frac{3}{2} k^{2}\lambda _{\rm D}^{2}
\end{equation}
can substantially alter the beam-plasma instability by changing the
wave-vector $\vec{k}$ along the trajectory of the beam. Here, $\vec{k}$ is
the wave-vector of the plasma wave, $n_{0}$ the background plasma density and
$\lambda_{\rm D}$ is the Debye length. The wave trajectory in the
inhomogeneous plasma can be described by the equations
\begin{eqnarray}
\frac{d\vec{r}}{dt}&=&\frac{\partial \omega }{\partial
\vec{k}} \\
\frac{d\vec{k}}{dt}&=&-\frac{\partial \omega }{\partial \vec{r}}
\end{eqnarray}
that results in a $\vec{k}$-vector change with the density that is described
by
\begin{equation}
\vec{k} \; \lambda _{\rm D}=\sqrt{\frac{2}{3\omega _{p}}\left[ \omega -\omega
_{p}\frac{\delta n(x)}{2n_0}\right] }
\end{equation}

The observations we present hereafter clearly show the simultaneous presence
of two oppositely propagating waves with $\vec{k}$-vectors close to each other,
so that $\vec{k}_1 \approx -\vec{k}_0$. Two scenarios can lead to such a
coexistence of primary and oppositely directed secondary waves. The first one
is the scattering off of density fluctuations, and the second one is
three-wave decay.

Let us consider the first scenario. Without loss of generality, we assume
here that the density increases along the x direction. In this case, the
$k_{\rm x}$ component of the wave-vector and the wave group velocity both
decrease along x. The wave amplitude then increases even in the absence of
growth or damping. This effect can be illustrated in the WKB approximation,
where the solution of the wave equation can be written in the form
\begin{equation}
E\sim E_{0}\sqrt{\frac{k_{0{\rm x}}}{k_{\rm x}\left( {\rm x}\right) }}\exp \left(
-j\omega t+j\int k_{\rm x}({x})\;dx\right) ,
\end{equation}%
and the amplitude increases when $k_{\rm x}\left({x}\right)$ decreases.
Without damping or growth, the component of the energy flux $F$ along the
density gradient can be supposed to be constant. We then have $F\sim V_{\rm
gr}E^{2}\sim k_{\rm x}E^{2}=$\,{const} and so the wave amplitude $E$
scales like $k_{\rm x}^{-1/2}$. In the WKB approximation, reflection
processes are supposed to be similar to mirroring, with part of the energy
being reflected and part damped. Two large amplitude waves, incident and
backscattered, can thus be observed simultaneously in the vicinity of the
reflection point \citep{bale00}. It is worth mentioning that the WKB
approximation is valid close to, but not at, the turning point where
$k_{\rm x}$ goes to zero. Immediately near this point the wave amplitude
resembles solutions of the Schr\"odinger equation with a linearly varying
potential expressed in terms of Airy functions and it remains finite at the
reflection point (see Sect.~17 in \citealp{ginzburg70}).

The second scenario is a change in the characteristics of the wave-wave
interaction by a variation of the wave-vector. The growth of primary waves
generated by beam-plasma interactions can be saturated by the decay process
$l_{0}\rightarrow l_{1}+s$, in which a primary Langmuir wave $l_{0}$ decays
into a secondary Langmuir wave $l_{1}$ and an ion-sound wave $s$. The
threshold condition for the decay instability is
\begin{equation}
\frac{\varepsilon _{0}E_{0}^{2}}{2n_{0} k_{B}T}>\frac{\nu _{\rm S}\nu _{\rm
L}}{ \omega _{\rm S}\omega _{\rm L}}.
\end{equation}
Here, $E_{0}$ is the amplitude of the primary wave, $T$ is the electron
temperature, $\nu _{\rm S}$ and $\nu _{\rm L}$ are respectively the damping
rates of the secondary ion sound and Langmuir waves, and $\omega _{\rm S}$
and $\omega _{\rm L}$ are their pulsations. When the primary wave propagates
into a denser plasma, the decrease in $k_{\rm x}$ leads to a decrease of
the wave-vector of the secondary waves. The phase velocity of the secondary
Langmuir wave then grows and the damping rate drops, finally initiating a
decay instability. It is well known from linear theory that the maximum
increment of decay instability then corresponds to almost exact back scatter
with a small correction to comply with momentum and energy conservation laws. As a
consequence, in this case the wave-vector $\vec{k}_{1}$ of the secondary
Langmuir wave becomes also approximately equal and opposite to that of the
primary wave, i.e.~$\vec{k}_{1}\approx -\vec{k}_{0}$.

These oppositely-directed waves provide a unique opportunity for measuring
their wavelength without requiring multipoint measurements.

\section{Expressions for evaluating the wavelength}
\label{sec_evaluation}

We assume that the frequency difference $\delta f=f_{0}-f_{1}$ (where $f_0$
and $f_1$ are respectively the frequencies of the incident and reflected
waves) is mainly caused by the Doppler shift rather than by the frequency
difference associated with the differing dispersions of the two waves.
Indeed, this frequency difference significantly exceeds that of the
ion-sound wave that might be generated by the decay process. In the
spacecraft reference frame, the wave frequencies are
\begin{equation}
f_{l}=f_{p}\left( 1+\frac{3}{2} k_{l}^{2}\lambda _{\rm D}^{2}\right)
+\frac{ \vec{k}_{l}\cdot \vec{V}_{\rm SW}}{2\pi }~~l=0,1
\end{equation}%
where ${V}_{\rm SW}$ is the velocity of the plasma with respect to satellite.
We assume that
\begin{equation}
\frac{\vec{k}_{l}\cdot \vec{V}_{\rm SW}}{\omega _{p}}\gg \frac{3}{2}
{k}_{l}^{2}\lambda _{\rm D}^{2} ,
\label{eq:dispersionless}
\end{equation}
and will check this {a posteriori}. Theoretical predictions and observations
both lend support to the alignment of the wave-vector of the primary (and
consequently secondary) waves along the magnetic field. Since the two
wave-vectors have almost equal magnitudes, we find
\begin{equation}
\delta f=f_{0}-f_{1}\simeq \frac{2\vec{k}_{0}\cdot \vec{V}_{\rm SW}}{2\pi }
=\frac{\vec{k}_{0}}{\pi }\left( \frac{\vec{B}}{B}\cdot \vec{V}_{\rm SW}\right),
\label{eq_themodel}
\end{equation}
where we suppose that both wave-vectors are approximately co-aligned with the
magnetic field. From this expression, we finally obtain the wavelength
$\lambda =2\pi /{k}_{0}$. Let us now show how these different quantities
are estimated.

%f2
\begin{figure}[t]
\vspace*{2mm}
\center
\includegraphics[width=9cm]{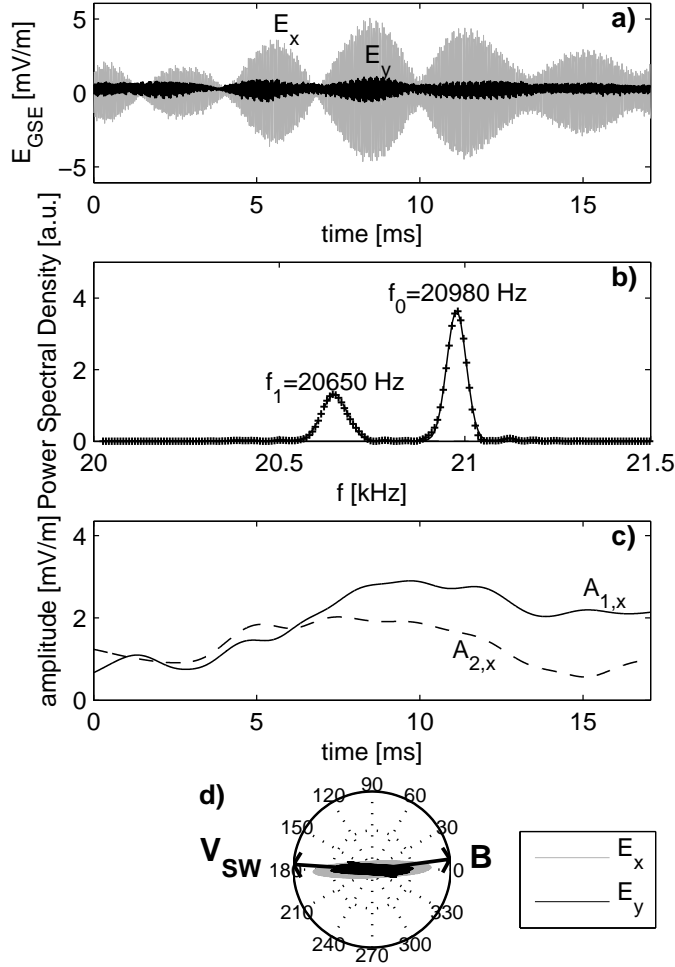}
\caption{Analysis procedure of a typical wave packet: {\bf(a)}~two components of
the electric field in GSE coordinates, {\bf(b)}~identification of the two peaks in
the power spectral density (continuous curves are gaussian fits),
{\bf(c)}~verification that the amplitudes and the phases (not shown) indeed vary
slowly in time, {\bf(d)}~hodogram to compare the polarisation with the orientation
of the magnetic field. The direction of the solar wind velocity varies
and is not necessarily aligned with the magnetic field.}
\label{fig_analysis}
\end{figure}

\section{Determining the wavelength in practice}
\label{sec_determining}

In practice, to measure the wavelength of the wave packets, we first need to
know their direction of propagation, which is a prerequisite for estimating
the wave-vector $\vec{k}$. We first convert the electric field components
into Geocentric Solar Ecliptic (GSE) coordinates, hence defining components
$E_{\rm x}$ and $E_{\rm y}$, see Fig.~\ref{fig_analysis}a. Next, a continuous Fourier
transform is applied in a small interval around the plasma frequency. Except
for some distorted wave-packets whose modulation width $T$ is only a few
oscillation periods, the power spectral density usually exhibits two distinct
peaks, see Fig.~\ref{fig_analysis}b. The relative width $\Delta f/f$ (where
$\Delta f$ is the full width at half maximum) of these peaks is always well
below 1\%, confirming the stability of the frequency over the interval. Each
component of the electric field can then be modeled as a sum of two complex
exponentials. A least-squares regression of the following model is performed
\begin{eqnarray*}
E_{i}(t) &=&A_{0,i}(t)\,{\rm Re}\left[ \exp {\left( j2\pi f_{0}t+j\phi
_{0,x}(t)\right) }\right] \\
&+&A_{1,i}(t)\,{\rm Re}\left[ \exp {\left( j2\pi f_{1}t+j\phi
_{1,i}(t)\right) }\right] \\
&+&C_{i}(t)+\epsilon _{i}(t)
\end{eqnarray*}
where $i=x$ or $y$ and $\epsilon (t)$ is the residual error to be minimized.
The amplitude $A(t)$ and the phase $\phi (t)$ vary slowly with respect to
$T$. Possible offsets are taken care of by a small and slowly varying
constant $C(t)$. We use a sliding gaussian window of width $W$ to estimate
these parameters at different parts of the wave-packet. Obviously, we must
have $1/f_{0}\ll W\ll T$. Taking $W=10/f_{0}$ was found to be a good
compromise. We checked that the amplitudes $A_{0}(t)$ and $A_{1}(t)$ show no
beating like the original wave packets do (Fig.~\ref{fig_analysis}c) and that
the phases $\phi _{0}(t)$ and $\phi_{1}(t)$ indeed vary slowly with respect
to the wave period.

Once the amplitudes and the phases are known, we can separate the components
of the electric field corresponding to the primary and secondary waves. A
plot in polar coordinates (see Fig.~\ref{fig_analysis}d) reveals their
polarisation in the xy-plane. Also shown are the projection on that plane
of the magnetic field $\vec{B}$ and the solar wind velocity $\vec{V}_{\rm SW}$. For
electrostatic waves, $\nabla \cdot \vec{E}=0\Rightarrow
\vec{k}\,/\!/\,\vec{E}$ and so the direction of the wave-vector $\vec{k}$
should be given by the semi-major axis of the polarisation ellipse.

Figure~\ref{fig_analysis}d shows two remarkable results, which are
systematically found for all the coherent wave-packets we analyzed. First,
both waves are almost linearly polarized and their polarisation ellipses have
the same orientation. We conclude that their wave-vectors are necessarily
parallel or antiparallel, as expected. Secondly, the angle between the
semi-major axis of the polarisation ellipse and the ambient magnetic field is
always small, rarely exceeding 20$\deg$. The absence of the third $E_{\rm z}$
component in principle prevents us from concluding that $\vec{k}\,/\!/\,\vec{B}$.
For all the events we analyzed, however, we always found the polarisation
ellipse to be parallel to the magnetic field, no matter how the latter is
oriented. We therefore conclude that both waves must necessarily propagate
along or very close to the magnetic field direction. Momentum conservation
finally implies $\vec{k}_{0}\approx - \vec{k}_{1}$.

%f3
\begin{figure}[t]
\vspace*{2mm}
\center
\includegraphics[width=9cm]{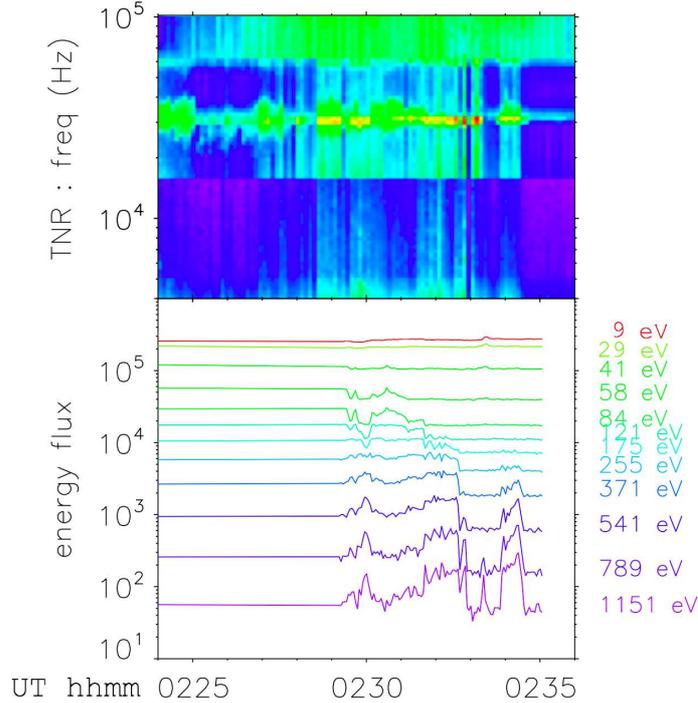}
\caption{Example of electric field spectrum from the WAVES instrument (top plot)
and electron flux from the Three-Dimensional Plasma and Energetic Particle Investigation
\citep{lin95} on 17 April 1996.}
\label{fig_electron_flux}
\end{figure}

For the example shown in Fig.~\ref{fig_analysis}, we measure $\delta
f=330$\,Hz, $f_{p}=20.815$\,kHz and so from Eq.~(\ref{eq_themodel}) we
finally obtain the wavelength $\lambda =2.57$\,km. We check that the waves
are indeed weakly dispersive, since from Eq.~(\ref{eq:dispersionless}) we
have
\begin{equation}
\frac{\vec{k}\cdot \vec{V}_{\rm SW}}{\omega _{p}}=0.017 \gg
\frac{3}{2}{k}^{2}r_{\rm D}^{2}=0.0031
\end{equation}
Similar values are obtained for the other coherent wave packets shown in
Fig.~1, except for the last one, for which $f_{0}$ and $f_{1}$ cannot be
properly identified. We analyzed 644 wave-packets that way and found the
wavelength to be on average $\lambda = 2.1$\,km, with a standard deviation of
0.5\,km; the typical Debye length in the solar wind at 1\,AU is 10\,m.

\section{Discussion and conclusion}

\label{sec_discussion}

The coherent wave packets we observe always coincide with an enhancement in
the energetic electron flux. The energy of these electrons ranges from
several hundreds of eV up to several keV. Figure~\ref{fig_electron_flux}
shows the energetic electron flux measured by WIND during a typical burst of
Langmuir wave activity. \citet{fitzenreiter96} have shown that these wave
packets are associated with \textit{bump on tail} features in the electron
distribution. Their wavelength then should be $\lambda =2\pi V_{\rm b}/\omega
_{p}$, where $V_{\rm b}$ is the beam velocity. For waves with a frequency of
20\,kHz that are generated by 1\,keV electron beams, we find $\lambda
=935$\,m. This value is about 2.3 times smaller than the one we measure. The
two values cannot be reconciled by assuming that the wave activity is
triggered by higher energy energy electrons, since unrealistic energies in
excess of 5\,keV would be required.

The wavelength we observe, however, is compatible with that found for wave
propagation in inhomogeneous plasma. As it was pointed out by
\citet{kellogg99} from the analysis of the density fluctuation spectra,
variations in the $\vec{k}$-vector, as caused by density fluctuations, can
exceed the vector magnitude. Therefore, the discrepancy we observe between
the predicted and observed wavelengths is most probably due to a change of
the wave-vector when the waves propagate into a higher density region.
Indeed, for the solar wind conditions we consider, the variation of the wave
dispersion can be as large as
\begin{equation}
\delta \omega /\omega _{p}=(3/2)k^{2}\lambda _{\rm D}^{2} \ \approx \
0.3\%.
\end{equation}
The associated level of density variations is
\begin{equation}
\delta n/n \ \approx \ 2\delta
\omega /\omega _{p}\ \approx \ 0.6\%.
\end{equation}
Such levels are compatible with estimates made at the electron foreshock \citep{neugebauer75, celnikier87, kellogg99}
and are large enough to offer an explanation for the relatively large
wavelengths we observe.

Our conclusion is further supported by recent direct observations by
\citet{ergun08} and \citet{krasno07} of unusually high levels of density
fluctuations in the solar wind. The conventional theory of beam-plasma
interaction that has been developed for homogeneous plasmas does not apply to
such cases. Indeed, our results show in an unambiguous way that the observed
wavelengths exceed their theoretical prediction. The theory should therefore
include the effect of large amplitude random density fluctuations from the
very beginning.

\subsection*{Acknowledgements}
{\small V.~K.~acknowledges CNES for financial support through Scientific Space
Research Proposal entitled ``Cluster Co-I DWP''.\\
\hspace*{4mm} Guest Editor M.~Gedalin thanks one anonymous referee for
her/his help in evaluating this paper.\\[2mm]

\includegraphics[width=7.4cm]{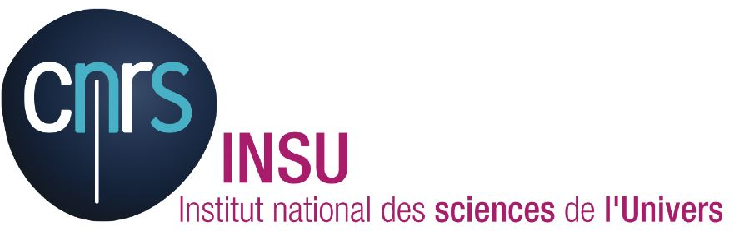}\\[3.5mm]
\mbox{The publication of this article is financed by CNRS-INSU.}
}

\end{document}